\documentstyle[aps,prl,twocolumn, epsf]{revtex}

\def\be{\begin{equation}}
\def\ee{\end{equation}}
\def\bea{\begin{eqnarray}}
\def\eea{\end{eqnarray}}
\def\bma{\begin{mathletters}}
\def\ema{\end{mathletters}}

\newcommand{\bra}[1]{\mbox{$\langle #1 |$}}
\newcommand{\ket}[1]{\mbox{$| #1 \rangle$}}

\newcommand{\proj}[1]{\ket{#1}\!\bra{#1}}

\begin{document}
\title{Optimal local preparation of an arbitrary mixed state of two qubits.\\
Closed expression for the single copy case.}
\author{Guifr\'e Vidal\cite{email}\\
Institut f\"{u}r Theoretische Physik,
                             Universit\"{a}t Innsbruck 
                             Technikerstra\ss e 25 
                            A-6020 Innsbruck, Austria.}
\maketitle

\begin{abstract}
In this note we consider the problem of preparing a {\em single} copy of an arbitrary two-qubit mixed state $\rho$ starting from an entangled pure state $\psi$ and using only local operations assisted with classical communication. We present an analytical expression for the minimal amount of pure state entanglement required, and describe the corresponding local strategy. We also examine optimal probabilistic generalizations of the previous process.

\end{abstract}
 
\section{Introduction.}

%
%

Quantum correlations are important not only as a fundamental aspect of quantum theory \cite{Ei} but also as a resource in recent developments of quantum information theory \cite{same}. Consequently, such correlations are presently subject to several investigations, both at the theoretical and experimental levels.


%
%
 pure-state entanglement has already been extensively studied for bipartite systems. As proposed in the pioneering works of Bennett et al. \cite{Be1}, the paradigmatic setting assumes that the distant parties sharing the composite system are only allowed to perform local operations and communicate classically (LOCC). Within this restricted set of transformations they will try to manipulate optimally the non-local resources contained in an initial entangled state. 
 This approach has been applied to two different, complementary contexts, for which all the relevant magnitudes have been successfully identified. And thus, explicit protocols for the optimal manipulation of pure-state entanglement are presently well-known.

 The so-called {\em finite regime} is concerned with the manipulation of a composite system with a finite dimensional Hilbert space. Arguably the main step forward was attained by Nielsen \cite{Ni}, who reported the conditions for a pure state $\psi$ to be locally convertible into another pure state $\phi$ deterministically. Subsequently, his result has been extended in several ways to the case when deterministic local transformations cannot achieve the target state $\phi$. Thus optimal conclusive \cite{Gui1}, probabilistic \cite{Jo1} and approximate \cite{Gui2} transformations are now well-known. Such efforts, that have also led to obtain closed expressions for entanglement concentration \cite{Ha,Jo1} and unveil the surprising phenomena of entanglement catalysis \cite{Jo2}, show that during entanglement manipulation some of the non-local properties of the system are irreversibly lost, and that entanglement does not behave in this regime as an additive quantum resource.

 Recall that a quantitative description of pure-state entanglement in a ${\cal C}^M\otimes{\cal C}^N$ bipartite system is given by a set of $n=$ min$(M,N)$ independent entanglement monotones \cite{Gui3} introduced by the author in \cite{Gui1}, in the sense that the local feasibility of a transformation is determined by the non-increasing character of all these functions. For instance, in a two-qubit system we have two monotones, namely 
\bea
E_1(\psi) &\equiv& \lambda_1+\lambda_2 = 1, \\
E_2(\psi) &\equiv& \lambda_2,
\eea
where $\lambda_1$ and $\lambda_2$ ($\lambda_1 \geq \lambda_2$) are the square of the two Schmidt coefficients of a pure state (or normalized vector) $\psi \in {\cal C}^2\otimes{\cal C}^2$. The parties can then conclusively transform the state $\psi$ into another state $\phi$ with an a priori probability of success $p$ {\em iff} \cite{Gui1}
\be
p \leq \min \{\frac{E_1(\psi)}{E_1(\phi)}, \frac{E_2(\psi)}{E_2(\phi)}\} = \min \{ 1, \frac{\lambda_2^{\psi}}{\lambda_2^{\phi}} \}.
\ee 
More generally, Jonathan and Plenio \cite{Jo1} showed that the parties can probabilistically convert the initial state into one of the states $\{\phi_i\}$, with a priori probability $p_i$ for outcome $i$, {\em iff} none of the monotones $E_k$ ($k=1,~..., n$) were increased on average during the transformation, i.e in the two-qubit case, {\em iff} $\sum_i p_i \leq 1$ and
\be
E_2(\psi) \geq \sum_i p_iE_2(\phi_i).
\label{jon}
\ee

 On the other hand the {\em asymptotic regime} --which was actually the first exhaustively studied \cite{Be1}-- assumes that the parties into play share an infinite number of copies of a given entangled state. It benefits from the possibility of using block-coding techniques and from the law of large numbers: it allows for some inaccuracy in the output of the transformation, which then becomes irrelevant in the limit of a large number of copies of the entangled state under manipulation. In this regime, the suitable one when dealing with asymptotic aspects of quantum information theory, the only relevant parameter is the entropy of entanglement ${\cal E}(\psi)$ \cite{Be1}, an additive measure which for two-qubit states reads
\be
{\cal E}(\psi) \equiv -\lambda_1 \log_2 \lambda_1 - \lambda_2 \log_2 \lambda_2.
\ee
 Its conservation dictates the ratio at which any two pure entangled states can be asymptotically converted into each other, and it implies that such conversions are fully reversible. Optimal local manipulation can be made asymptotically with a vanishing classical communication cost \cite{Lo}, which confirms entanglement as a truly inter-convertible resource.


%
%

 Despite its behavior being presently quite well understood, pure-state entanglement is just an idealization. In any realistic physical situation noise plays its role and the state of the system is unavoidably mixed. Thus, understanding also mixed-state entanglement is necessary in order to be able, in practice, to successfully exploit this quantum resource.

 From a theoretical point of view, entangled mixed states turn out to be more difficult to deal with. To begin with, no practical criterion is known that tells us in general whether a mixed state is entangled or separable (unentangled) \cite{HLVC}. Then, although it has been remarkably shown that some entangled mixed states can be asymptotically distilled into pure entangled states \cite{Be2} (that is, they contain {\em distillable entanglement}), also that others cannot \cite{Ho} (they only contain {\em bound entanglement}), it is again not even known how to identify in general these states. At a quantitative level, the amount of pure-state entanglement needed to prepare a given mixed state (or its {\em entanglement of formation} \cite{Be1}) and that which can be distilled out of it (its {\em distillable entanglement} \cite{Be2}) --both {\em asymptotic} measures-- remain to be computed, a celebrated exception being Wootters' closed expression for the entanglement of formation ${\cal E}(\rho)$ of an arbitrary state of two qubits \footnote{Although the interpretation of Wootters' expression as the entanglement of formation relies on the unproved assumption that such a measure of entanglement is additive.} \cite{Wo}.

 In this note we present the {\em finite-regime} analog of Wootters' results, namely a closed expression for the minimal amount of pure-state entanglement required in order to create a single copy of an arbitrary mixed state of two qubits. This expression will allow us to determine, for instance, whether the parties can locally create a given mixed state $\rho$ starting from some pure entangled state $\psi$. Also, in those cases where this is not possible with certainty, we will be able to construct a local strategy which succeeds with the greatest a priori probability, and, more in general, we will be able to assess the feasibility by means of LOCC of the most general transformation starting from a pure state $\psi$, namely that producing one of the final states $\{\rho_i\}$ with corresponding a priori probabilities $\{p_i\}$.

It turns out that the parameter governing all the transformations above is an extension to mixed states of the entanglement monotone $E_2$. Recall that two-qubit pure states depend only on one independent non-local parameter (for instance the smallest Schmidt coefficient) and thus it is not surprising that $E_2$ suffices to quantify their entanglement. More remarkable is the fact that just one extension of the same parameter also rules the local preparation procedures for mixed states, the set of mixed states depending on $9$ non-local parameters \cite{comment}.

\section{Closed expression for $E_2(\rho)$ in a two-qubit system.}

 One of the main problems met while quantifying entanglement of mixed states in terms of pure-state entanglement comes from the fact that any mixed state $\rho$ accepts many decompositions as a mixture of pure states, namely
\be
\rho= \sum_k p_k \proj{\psi_k}
\label{decomp}
\ee
for infinitely many pure-state ensembles $\{\psi_k,p_k\}$.
 Let us consider a generic entanglement monotone $\mu(\psi)$ for pure states (see \cite{Gui3} for a general way to construct  $\mu(\psi)$). It turns out to be often interesting to extend it to mixed states as a convex roof --which preserves its monotonicity under LOCC-- by defining
\be
\mu(\rho) \equiv \min_{\{\psi_k,p_k\}} \sum_k p_k \mu(\psi_k),
\ee
where the minimization needs to be performed over all pure-state ensembles $\{\psi_k,p_k\}$ realizing $\rho$ as in (\ref{decomp}). This is typically a strenuous optimization problem that prevents from obtaining an analytical expression for any of such measures. The value of $\mu(\rho)$ must then rely on impractical numerical computations.

 In spite of being a difficult task, Wootters \cite{Wo} did solve analytically this optimization for the particular case of $\mu(\psi) = {\cal E}(\psi)$ (entropy of entanglement) and a two-qubit system (${\cal H} = {\cal C}^2\otimes{\cal C}^2$). Next we consider and solve, also for the two-qubit case, the choice $\mu(\psi) = E_2(\psi)$. We start by briefly rephrasing Wootters' argumentation, of which we will be making a substantial use.


 Wootters' strategy consisted in 
\begin{itemize}
\item introducing the so-called concurrence, defined for two-qubit pure states as $C(\psi) \equiv 2\sqrt{(1-\lambda_2)\lambda_2}$ and extended to mixed states as
\be
C(\rho) \equiv \min_{\{\psi_k,p_k\}} \sum_k p_k C(\psi_k);
\label{defcon}
\ee
\item computing $C(\rho)$ for any two-qubit mixed state; 

\item showing that the convex roof ${\cal E}(\rho)$ of the entropy of entanglement ${\cal E}(\psi)$ increases monotonically with ${\cal C}(\rho)$.
\end{itemize}

 Let us recover the closed expression for the concurrence: denote by $\bar{\rho}$ the complex conjugation of an arbitrary two-qubit mixed state $\rho$ in the standard local basis $\{\ket{00},\ket{01},\ket{10},\ket{11}\}$, and by $\sigma_y$ the matrix $\left( \begin{array}{cc} 0 & -i\\ i & 0 \end{array}\right)$. The ``spin-flipped'' density matrix $\tilde{\rho}$ is defined as $(\sigma_y \otimes \sigma_y)\bar{\rho}(\sigma_y \otimes \sigma_y)$. Then Wootters proved  \cite{Wo} that the concurrence of $\rho$ is given by 
\be
C(\rho) = $ max $\{v_1-v_2-v_3-v_4, 0\},
\label{closed}
\ee
$v_i$ being the square roots of the eigenvalues of $\rho\tilde{\rho}$, in decreasing order. In the way he also proved that for any mixed state $\rho$ there is always an optimal pure-state ensemble with at most four states, and with all of them having the same concurrence, i.e.
\be
\rho = \sum_{k=1}^{l\leq 4} p_k \proj{\phi_k}, ~~~~C(\rho)=C(\phi_k).
\label{mini}
\ee 

The closed expression (\ref{closed}) will be very useful for the purposes of this note, because as we will show next also the entanglement monotone $E_2$ for mixed states can be expressed in terms of the concurrence. Explicitly,

\vspace{2mm}

{\bf Result:}
\be
E_2 (\rho) = \frac{1-\sqrt{1-C(\rho)^2}}{2}.
\label{mixe2}
\ee

{\bf Proof:}
It is essential to notice that the function $E_2(\psi) = \lambda_2$ is a convex, monotonically increasing function of $C(\psi)$ for pure states (which is exactly what happens also with the entropy of entanglement, the following argument being parallel to that implicit in \cite{Wo}). Thus, we have that 
\be
E_2(\psi) = \frac{1-\sqrt{1-C(\psi)^2}}{2}.
\ee
We first want to see that the value of
\be
E_2(\rho) \equiv \min_{\{\psi_k,p_k\}} \sum_k p_k E_2(\psi_k)
\label{definition2}
\ee
cannot be smaller than that of equation (\ref{mixe2}). But let us just suppose that this were not the case, i.e. that we could find a pure-state ensemble $\{\dot{\psi}_k, \dot{p}_k\}$ for $\rho$ such that 
\be
\sum_k \dot{p}_k E_2(\dot{\psi}_k) < \frac{1-\sqrt{1-{\cal C}(\rho)^2}}{2}.
\label{ineq}
\ee
 Define the function $f(x)\equiv 2\sqrt{(1-x)x}$, which in the relevant interval $x \in [0,1/2]$ is a concave, monotonically increasing function. Then it would follow, by taking both sides of equation (\ref{ineq}) as arguments of this function and using its concavity [namely $\sum_k p_kf(x_k) \leq f(\sum_k p_kx_k)$], that $\sum_k \tilde{p}_k {\cal C}(\tilde{\psi}_k) < {\cal C}(\rho)$, which would be in contradiction with the definition (\ref{defcon}). Finally, the four-term, pure-state ensemble (\ref{mini}) achieves the optimal value (\ref{mixe2}) for $E_2$. $\Box$

 \section{Optimal preparation of a two-qubit mixed state.}

 As we will explain now, the explicit expression (\ref{mixe2}) for $E_2(\rho)$ is useful because it tells us explicitly whether the entanglement of a pure state $\psi$ allows us to locally create the state $\rho$. We can then construct local preparation procedures for mixed states that require the minimal amount of entanglement.

 In \cite{Gui2} we discussed the necessary and sufficient conditions that make possible, by means of LOCC, a probabilistic transformation $\psi \rightarrow \{\rho_i, p_i\}$, namely that of the pure state $\psi$ into one of the mixed states $\{\rho_i\}$ with corresponding a priori probabilities $\{p_i\}$. Notice that this is the most general transformation a pure state can undertake. The existence of the inequivalent pure-state monotones $E_2, ..., ~E_{\min(M,N)}$ \cite{Gui1} in a generic ${\cal C}^M\otimes{\cal C}^N$ system did not allow to express such conditions in terms of their convex roof extensions $E_2(\rho_i), ~...,~ E_{\min (M,N)}(\rho_i)$. However, for a ${\cal C}^2\otimes{\cal C}^2$ system, and actually also for any ${\cal C}^2\otimes{\cal C}^N$ system, we announce:

{\bf Theorem 1.a:} The transformation $\psi \rightarrow \rho$ can be achieved by means of LOCC {\em iff}
\be
E_2(\psi) \geq E_2(\rho),
\label{condition1}
\ee
where $E_2(\rho)$ corresponds to the convex roof extension of the entanglement monotone $E_2(\phi) \equiv \lambda_2^{\phi}$ as in (\ref{definition2}).

\vspace{2mm}

 Notice that when condition (\ref{condition1}) is fulfilled, then we can say that the pure state $\psi$ is more entangled than the mixed state $\rho$, this being an incomplete extension to mixed states of Nielsen's partial order for pure states \cite{Ni}. 

More generally, we can announce

\vspace{2mm}

{\bf Theorem 1.b:} The transformation $\psi \rightarrow \{\rho_i, p_i\}$ can be achieved by means of LOCC {\em iff}
\be
E_2(\psi) \geq \sum_i p_iE_2(\rho_i).
\label{condition1b}
\ee

\vspace{2mm}

{\bf Corollary:} The maximal a priori probability $P(\psi \rightarrow \rho)$ of successfully transforming $\psi$ into $\rho$ by means of LOCC is given by 
\be
P(\psi \rightarrow \rho) = \min \{\frac{E_1(\psi)}{E_1(\rho)}, \frac{E_2(\psi)}{E_2(\rho)} \} = \min \{ 1, \frac{\lambda_2^{\psi}}{E_2(\rho)} \}.
\label{coro}
\ee
 
 We remark that the previous statements, when complemented with expression (\ref{mixe2}) for the two-qubit case, result in a complete and explicit account of what the set of transformations so-called LOCC can produce out of the initial pure state $\psi$.

 Let us marginally note that had the parties started with a pure state $\psi$ from a ${\cal C}^M\otimes{\cal C}^N$ system, $M,N > 2$, the previous results would still hold, but with $E_2(\phi) \equiv \sum_{i\geq 2} \lambda_i^{\phi} = 1 - \lambda_1^{\phi}$ (here, as before, $\lambda_i^{\phi}$ are the square of the Schmidt coefficients of $\phi$, ordered decreasingly).

 Now, the results above follow straightforwardly from the fact that the entanglement monotone $E_2$ does not increase on average under LOCC, and from the fact that for any transformation such that $E_2$ is not increased we can find an explicit local protocol realizing it. Indeed, to see the latter let us suppose, for instance, that Alice and Bob want to prepare locally the state $\rho$ from $\psi$ and that condition (\ref{condition1}) is fulfilled. This means that we can think of $\rho$ as a probabilistic mixture $\sum_k p_k \proj{\psi_k}$, where the pure-state ensemble satisfies that $E_2(\psi) \leq \sum_k p_k E_2(\psi_k)$. That is, condition (\ref{jon}) is fulfilled and the probabilistic transformation $\psi \rightarrow \{\psi_k, p_k\}$ can be realized locally (see \cite{Jo1} for an explicit protocol). After the probabilistic transformation the parties only need to discard the information concerning the index $k$ in order to obtain the state $\rho$. The generalization to a probabilistic transformation $\psi \rightarrow \{\rho_i, p_i\}$ is straightforward, as we only need to add an extra index $i$ to the previous protocol, which is not discarded in the final step.

 As mentioned before, Wootters \cite{Wo} also proved that in the ${\cal C}^2\otimes{\cal C}^2$ case we can always find an optimal decomposition (\ref{mini}) for $\rho$ such that there are at most only four states, they being equally entangled, and thus also $\lambda_2^{\phi_k} = E_2(\rho)$ for $k=1,~..., l;~ l\leq 4$. This provides us with somehow simple protocols. For instance, we can construct a local, optimal preparation procedure for $\rho$ by composing the following two steps: first, Alice and Bob transform $\psi$ into one of the states $\phi_k$, say $\phi_1$, following Nielsen's deterministic protocol \cite{Ni}; then they choose, randomly and with a priori probabilities $\{p_k\}$, to perform one of the bi-local unitary operations $\{U_k\otimes V_k\}$, where $\phi_k = U_k\otimes V_k \phi_1$. Finally, they discard the information concerning the index $k$. Similar procedures also apply for the probabilistic and conclusive transformations (cf. equations (\ref{condition1b}) and (\ref{coro})).

\section{conclusions}

We have analyzed the problem of optimally preparing a two-qubit mixed state by means of LOCC, when the parties initially share an entangled pure state. We have presented necessary and sufficient conditions for such preparation to be possible in terms of the entanglement monotone $E_2(\rho)$, for which we have obtained a closed expression. These results highlight the role the quantity $E_2$ plays in ${\cal C}^2\otimes{\cal C}^N$ systems: not only does it determines whether a pure-state transformation can be performed locally, but it also provides the preparation cost for mixed states.

In view of the fact that almost no closed expression for mixed state entanglement is known, we expect our result to be also of practical interest as a handy tool for future related studies.

\vspace{2mm}

The author thanks Wolfgang D\"ur and Ignacio Cirac for their comments. He also acknowledges a CIRIT grant 1997FI-00068 PG (Autonomic Government of Catalunya) and a Marie Curie Fellowship HPMF-CT-1999-00200 (European Community).


\begin{references}
\bibitem{comment} A mixed state of two qubits is a trace-normalized, $4\times 4$ hermitian matrix, and thus depends on $15$ parameters. We subtract from them the $6$ relevant degrees of freedom corresponding to local unitaries to obtain the $9$ non-local parameters.
\end{references}
\end{document}